\documentclass[twocolumn,showpacs,groupedaddress]{revtex4}

\usepackage{graphicx}
\begin{document}
\title{Gauge field in ultra-cold bipartite atoms}
\author{H. Wang, W. Wang, and X. X. Yi \footnote{Email:yixx@dlut.edu.cn}}
\affiliation{Department of Physics, Dalian University of
Technology, Dalian 116024, China}
\date{\today}
\begin{abstract}
The effects of entanglement and spin-spin collision on the gauge
field in ultracold atoms are presented in this paper. Two gauge
fields are calculated and discussed. One of the fields comes from
space dependent spin-spin collisions in ultra-cold atoms, while
another results from the usual Born-Oppenheimer method, which
separates the center-of-mass motion from the relative motion in
the two-body problem. Adiabatic conditions that lead to the key
results of this paper are also presented and discussed.
Entanglement shared between the two atoms is shown to affect the
atomic motion. In the presence of entanglement, the additional
scalar potential disappears, this is different from the case of
atoms in separable states.
\end{abstract}
 \maketitle

Gauge potentials have been found to appear very naturally in the
description of quantum mechanical systems, which depend upon
slowly varying external parameters. Beside the central role the
gauge fields play in the theory of fundamental interactions, gauge
fields are of interest in a variety of single- and many-body
quantum systems, leading to a variety of phenomena, for example
quantum Hall effect \cite{klitzing80, tsui82}. In recent years
ultra-cold atomic gases\cite{pitaevskii03} have become an ideal
playground to experimentally investigate many-body physics. This
is due to the advances in experimental techniques available in
atomic and optical physics. Recently several experimental groups
have succeeded in trapping and cooling fermionic
atoms\cite{demarco99,hadzibabic03} well below the Fermi
temperature. This is fascinating as it provides us an interface
between ultra-cold atoms and fermions in solid state systems.
Fermi systems are well known from the study of electron properties
in materials, however trapped atomic fermions are electrically
neutral and each atom has internal structure, hence directly
mapping from the electron properties into the atomic one is not
necessarily straightforward.

Combining the ultra-cold atoms and gauge field, a interesting
study is to create an effective magnetic filed for ultra-cold
atoms\cite{juzeliunas04}, the method is based on light-induced
gauge potentials for atom with a space-dependent dark
state\cite{harris97}. The dark state is created in three-level
$\Lambda$-type atoms interacting with two laser fields  under
conditions of electromagnetically induced
transparency\cite{lukin03}. A vector gauge potentials arises for
the adiabatic center-of-mass motion, as the dark state is space
dependent. From the side of trapped atoms, the strength of the
atomic pair interaction can be strong relying on the magnetically
tuned Feshbach resonance\cite{bryant77}. These give rise to a
question of how the gauge fields depend on the inter-couplings
between the two Fermi atoms. Alternatively, it is believed that
intersubsystem couplings would lead to entanglement, then how the
entanglement affects the gauge field? In this paper, we will try
to answer these questions by studying the gauge field for
ultra-cold atoms with spin-dependent collisions and investigating
the effect of entanglement on the gauge field. This is of
relevance to the recent study on BCS-BEC crossover in ultra-cold
$^{40}K$ gases\cite{parish05}, where the inter-atomic interaction
is specified by a two-body potential that depends only on the
electron spin
\begin{equation}
V(\vec{r}_1-\vec{r}_2)=V_c(\vec{r}_1-\vec{r}_2)+V_s(\vec{r}_1-\vec{r}_2)
\vec{S_1}\cdot\vec{S_2}.
\end{equation}
Here the spin states are $|F, m_F\rangle =\{ |\frac 9 2 ,-\frac 9
2 \rangle ,|\frac 9 2 ,-\frac 7 2 \rangle\}$ for $^{40}K$. This
collision preserves the total spin projection of the two-body
system, and then any scattering process between atomic states that
conserves the total spin projection is allowed. Taking the atomic
motion $\vec{p}_i^2/2m_i (i=1,2)$ and trapping potentials
$V_i(\vec{r}_i)$ into account, the total Hamiltonian of the system
reads,
\begin{eqnarray}
H&=&\frac{\vec{p}_1^2}{2m_1}+\frac{\vec{p}_2^2}{2m_2}+V_1(\vec{r}_1)
+V_2(\vec{r}_2)+V(\vec{r}_1-\vec{r}_2)+H_0,\nonumber\\
H_0&=&\frac{\hbar\omega_1}{2}\sigma_{1z}+\frac{\hbar\omega_2}{2}\sigma_{2z},
\end{eqnarray}
where $\sigma_{iz} (i=1,2)$ stand for the Pauli matrix, and
$\omega_i$ is the Rabi frequencies. As you will see, the
spin-dependent collision
$V_s(\vec{r}_1-\vec{r}_2)\vec{S_1}\cdot\vec{S_2}$ together with
the energy difference $(\hbar\omega_2-\hbar\omega_1)$ lead to a
gauge field $\vec{A}$, which would affect the relative motion of
the two atoms. This is different from the earlier study in
diatom\cite{moody86}. For fixed $(\vec{r}_1-\vec{r}_2)$ the
Hamiltonian
$H_s=H_0+V_s(\vec{r}_1-\vec{r}_2)\vec{S_1}\cdot\vec{S_2}$ can be
diagonalized to give a set of eigenvactors $|E_i\rangle
(i=1,2,...,4)$ and corresponding  eigenvalues $E_i$ as follows,
\begin{eqnarray}
|E_1\rangle&=&|\downarrow\downarrow\rangle,
E_1=\frac{-\hbar\omega_1-\hbar\omega_2}{2}+V_s\nonumber\\
|E_2\rangle&=&|+\rangle ,
E_2=-V_s+\sqrt{(\frac{\hbar\delta}{2})^2+V_s^2}\nonumber\\
|E_3\rangle&=&|-\rangle ,
E_3=-V_s-\sqrt{(\frac{\hbar\delta}{2})^2+V_s^2}\nonumber\\
|E_4\rangle&=&|\uparrow\uparrow\rangle ,
E_4=\frac{\hbar\omega_1+\hbar\omega_2}{2}+V_s.\nonumber\\
\end{eqnarray}
Here,
\begin{eqnarray}
\cos\theta=\frac{\frac{\hbar\delta}{2}}
{\sqrt{(\frac{\hbar\delta}{2})^2+V_s^2}},
\end{eqnarray}
$\hbar\delta=\hbar\omega_2-\hbar\omega_1$ and
\begin{eqnarray}
|+\rangle&=&-\cos\frac{\theta}{2}|\downarrow\uparrow\rangle
-\sin\frac{\theta}{2}|\uparrow\downarrow\rangle,\nonumber \\
|-\rangle&=&-\sin\frac{\theta}{2}|\downarrow\uparrow\rangle
+\cos\frac{\theta}{2}|\uparrow\downarrow\rangle.
\end{eqnarray}
By introducing $
\vec{R}=\frac{m_1\vec{r_1}+m_2\vec{r_2}}{m_1+m_2}$,
$\vec{r}=\vec{r_1}-\vec{r_2},$ $\mu=\frac{m_1m_2}{m_1+m_2}$,
$M=m_1+m_2,$   $\vec{P}=\vec{p_1}+\vec{p_2},$ and
$\vec{p}=\vec{p_1}-\vec{p_2},$ we can expand the full system
wavefunction $|\phi(R,r)\rangle$ as
\begin{equation}
|\phi(\vec{R},\vec{r})\rangle
=\sum_{n=1}^4\psi_n(\vec{R},\vec{r})|E_n(\vec{r})\rangle,
\label{expan}
\end{equation}
where $\psi_n$ is a composite wavefunction that describes the
relative and translational motion of the two atoms in the internal
state $|E_n(r)\rangle.$ Substituting equation Eq.(\ref{expan})
into Schr\"odinger equation, we get a set of coupled equations for
the components $\psi_n(\vec{R},\vec{r})$. These equations can be
written in a simple form by introducing column
$\psi=(\psi_1,\psi_2,\psi_3,\psi_4)^\top$ as
\begin{eqnarray}
i\hbar\frac{\partial}{\partial
t}\psi=[\frac{1}{2\mu}(-i\hbar\nabla-\vec{A})^2+V+\frac{\vec{P}^2}{2M}]\psi,
\label{scheq1}
\end{eqnarray}
where $\vec{A}$ and $V$ are 4 by 4 matrices given by
\begin{eqnarray}
\vec{A}_{mn}=i\hbar\langle E_m(\vec{r})|\nabla
E_n(\vec{r})\rangle,
\end{eqnarray}
and
\begin{eqnarray}
& &V_{mn}(\vec{R},\vec{r})=E_m(\vec{r})\delta_{mn}\nonumber\\
 &+&\langle
E_m(\vec{r})|V_c(\vec{r}) +V_1(\vec{R},\vec{r})
+V_2(\vec{R},\vec{r})|E_n(\vec{r})\rangle.
\end{eqnarray}
Clearly, matrix $\vec{A}$ includes contributions only from
internal atomic degrees, while $V$ includes contributions from
both the internal and external degrees of atoms. Thus $\vec{A}$
changes the relative motion of the two atoms, whereas $V$ affects
both the relative and translational motion of the system. Let us
now calculate and discuss the vector gauge field $\vec{A}$ in
detail. By the expression of  the matrix $\vec{A}$, it is easy to
show that the elements of $\vec{A}$ take the following form,
\begin{eqnarray}
A_{22}&=&A_{33}=0, \nonumber\\
A_{23}&=&i\hbar\langle +|\nabla
|-\rangle=\frac{1}{2}i\hbar\nabla\theta,\nonumber\\
A_{32}&=&i\hbar\langle -|\nabla
|+\rangle=-\frac{1}{2}i\hbar\nabla\theta,
\nonumber\\
\mbox{others}&=&0.
\end{eqnarray}
This leads to $\vec{B}=\nabla\times\vec{A}=0$, i.e., the vector
gauge field yields a vanishing magnetic field.   Consider two
limiting cases $|\hbar\delta|\gg 2|V_s|$ and $|\hbar\delta|\ll
2|V_s|,$ the first corresponds to a very small internal energy
difference between the atoms, with respect to the collision
strength $V_s$. In this situation $\cos\theta\simeq
1-V_s/(\hbar\delta)$ and $\sin\theta \simeq 2V_s/(\hbar\delta)$ up
to  first order in $|V_s/\hbar\delta|$. It yields
$\nabla\theta\simeq 2\nabla V_s/(\hbar \delta)$. Similar analysis
leads to $\nabla \theta\simeq \frac{\hbar\delta}{2V_s}\frac{\nabla
V_s}{V_s},$ for $ |\hbar\delta|\ll  2 |V_s|.$ With all these
together one arrives at
\begin{eqnarray}
\vec{A}_{23}&=& \frac{i\nabla V_s}{\delta}, \mbox{for}\
\ \hbar|\delta|\gg  2 |V_s|, \nonumber\\
\vec{A}_{23}&=&\frac{i\hbar^2\delta \nabla V_s}{4V_s^2},
\mbox{for}\ \  \hbar|\delta|\ll  2 |V_s|.\label{aapp}
\end{eqnarray}
Eq.(\ref{aapp}) shows that, the non-zero elements of matrix
$\vec{A}$ are proportional to $\nabla V_s$, the gradient of the
collision strength. Suppose $V_s\sim -1/r^n$, one gets
$\vec{A}_{23}\sim 1/r^{n+1}$ in the limit of $\hbar|\delta|\gg 2
|V_s|$, whereas $\vec{A}_{23}\sim r^{n-1}$ in $\hbar|\delta|\ll 2
|V_s|$ limit. The dependence of the imaginary part of
$\vec{A}_{23}$ on the energy difference $\delta$ as well as $r$
was shown in figure 1. For plotting figure 1, we chose
$V_s=-c_6/r^6$, which is the leading part of atom-atom collisions
in ultra-cold Alkali atoms\cite{vogels97}.
\begin{figure}
\includegraphics*[width=0.7\columnwidth,
height=0.6\columnwidth]{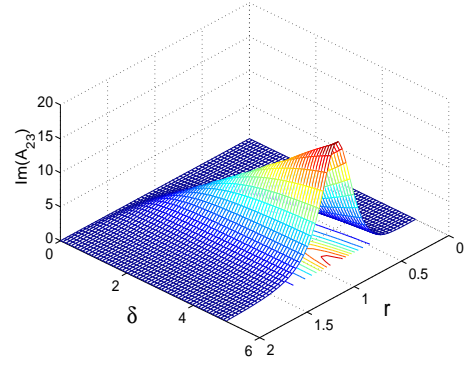} \caption{ Imaginary part of
$\vec{A}_{23}$ {\it versus} $\delta$ and the relative distance
$r$. $\hbar=1$ and $c_6=1$ were chosen for this plot. }
\label{fig1}
\end{figure}

Eq.(\ref{scheq1}) is analytically exact, which describes the joint
dynamics of the relative and translational  degrees of freedom.
Suppose that the atomic state $|+\rangle$ is well separated from
the remaining atomic states $|E_1\rangle$, $|E_3\rangle$ and
$|E_4\rangle$. Neglecting transitions from $|E_2\rangle$ to the
others, we get an effective equation for the atomic motion in the
internal state $|E_2\rangle$.
\begin{equation}
i\hbar\frac{\partial}{\partial
t}\psi_2=[\frac{1}{2\mu}(-i\hbar\nabla)^2+
U+\frac{\vec{P}^2}{2M}]\psi_2,\label{scheq2}
\end{equation}
where $U=V_{22}+\frac{1}{2\mu}\vec{A}_{23}\cdot \vec{A}_{32}.$ An
additional scalar potential $1/(2\mu)\vec{A}_{23}\cdot
\vec{A}_{32}$ appears due to the exclusion of the internal states
$|E_1\rangle$, $|E_3\rangle$ and $|E_4\rangle$ in the effective
equation of motion Eq.(\ref{scheq2}). The physics behind this may
be understood as adiabatic eliminations of the internal states
$|E_1\rangle$, $|E_3\rangle$ and $|E_4\rangle$, resulting in an
effective potential like stark shifts. This result is interesting
because it provides us another way to test the effect of the
vector potential, which yields zero magnetic fields $\vec{B}.$ The
Schr\"odinger equation which governs the adiabatic evolution of
atoms in state $|-\rangle$ is similar to that for atoms in state
$|+\rangle$. The difference is only in the effective potential
$U$, in this case it is $U=V_{33}+\frac{1}{2\mu}\vec{A}_{23}\cdot
\vec{A}_{32}.$
 In the Born-Oppenheimer
method\cite{born27}, eigenfunction $\psi_2$ is decomposed into
relative components $\psi^r_2(\vec{R},\vec{r})$ and components of
center-of-motion $\Phi^c(\vec{R})$
\begin{equation}
\psi_2(\vec{R},\vec{r})=\sum_{\alpha}\Phi^c_{\alpha}(\vec{R})
\psi_{2,\alpha}^r(\vec{R},\vec{r}),
\end{equation}
The $\psi_{2,\alpha}^r$ form a basis of eigenfunctions for the
relative motion Hamiltonian
$\frac{1}{2\mu}(-i\hbar\nabla)^2+U(R,r)$ for fixed $\vec{R}$ when
the translational kinetic energy term $\vec{P}^2/2M$ is ignored,
namely
\begin{eqnarray}
[\frac{1}{2\mu}(-i\hbar\nabla)^2+U(\vec{R},r)]
\psi^r_{2,\alpha}(\vec{R},\vec{r})=\nonumber\\
E_r^{\alpha}(\vec{R})\psi^r_{2,\alpha}(\vec{R},\vec{r}).
\end{eqnarray}
The vector wave functions
$\Phi^c(\vec{R})=(\Phi^c_1(\vec{R}),...,\Phi^c_N(\vec{R}))^\top$
then satisfy
\begin{eqnarray}
[-\frac{\hbar^2}{2M}(\nabla_R-i\vec{A}^R)^2+E_r^{\alpha}(\vec{R})]
\Phi^c_{\alpha}(\vec{R})=E_T\Phi^c_{\alpha}(\vec{R}),\label{tra1}
\end{eqnarray}
where
\begin{equation}
\vec{A}^R_{\alpha \alpha}=i  \langle
\psi_{2,\alpha}^r|\nabla_R|\psi_{2,\alpha}^r\rangle,
\end{equation}
with $\langle
\vec{R},\vec{r}|\psi_{2,\alpha}^r\rangle=\psi_{2,\alpha}^r$ To
write Eq.(\ref{tra1}), we assume that the two atoms remain in the
$\alpha$  level, requiring that the translational motion vary very
slowly with respect to the relative motion. In the case when this
term could not be ignored, the Hamiltonian which governs the
evolution of the vector wave functions $\Phi^c(\vec{R})$ is
\begin{eqnarray}
H_{\alpha \beta}=-\frac{\hbar^2}{2M}\sum_{\gamma}\left [
\nabla_R+\langle
\psi^r_{2,\alpha}|\nabla_R|\psi_{2,\gamma}^r\rangle\right ]
\cdot\nonumber\\
\left [ \nabla_R+\langle
\psi^r_{2,\gamma}|\nabla_R|\psi_{2,\beta}^r\rangle\right ] +
E_r^{\alpha}(\vec{R})\delta_{\alpha \beta},
\end{eqnarray}
the off-diagonal terms in $H_{\alpha \beta}$ would result in
transitions between states $|\psi_{2,\alpha}^r\rangle$ with
different $\alpha$. Eq.(\ref{tra1}) holds when the adiabatic
condition satisfied. Mathematically, this condition may be
expressed as  $(\alpha\neq \beta)$,
\begin{equation}
\left|\frac{\langle\psi_{2,\alpha}^r|H_{\alpha\beta}|\psi_{2,\beta}^r\rangle}
{E_r^{\alpha}-E_r^{\beta}}\right |\ll 1,
\end{equation}
i.e., the transition induced by the off-diagonal terms in
$H_{\alpha\beta}$ may be ignored.

Now we turn to study the effect of entanglement on the gauge
fields in ultracold atoms. Suppose that the bipartite atoms are in
entangled states
\begin{equation}
|\psi\rangle=\sum_n\Phi_n(\vec{R},\vec{r},t)|\alpha_n(\vec{R})\rangle_a\otimes
|\beta_n(\vec{r})\rangle_b,\label{ens1}
\end{equation}
where $|\alpha_n(\vec{R})\rangle_a$ ( $
|\beta_n(\vec{r})\rangle_b$)$(n=1,2,...,N)$ represent internal
states of atom a (atom b). The entanglement shared between the
atoms may be measured by the quantum entropy as
$s=-\sum_n|\Phi_n(\vec{R},\vec{r},t)|^2 ln
|\Phi_n(\vec{R},\vec{r},t)|^2,$ which would remain unchanged if
there are no couplings between them. Consider a Hamiltonian
\begin{equation}
H=\frac{\vec{P}_a^2}{2M}+\frac{\vec{p}_b^2}{2m}+H_a(\vec{R})+H_b(\vec{r}),
\end{equation}
which governs the evolution of the bipartite atoms without
couplings between them. Here, $H_a(\vec{R})$ ($H_b(\vec{r})$)
denotes the Hamiltonian of the electronic degrees of freedom of
atom a (atom b), and satisfies
\begin{equation}
H_a(\vec{R})|\alpha_n(\vec{R})\rangle_a=\varepsilon_n(\vec{R})
|\alpha_n(\vec{R})\rangle_a.\label{res1}
\end{equation}
Similarly, we require that
\begin{equation}
H_b(\vec{r})|\beta_n(\vec{r})\rangle_b=\epsilon_n(\vec{r})
|\beta_n(\vec{r})\rangle_b.\label{res2}
\end{equation}
These restrictions can be relieved and we will show later that
they do not change the results. Substituting Eq.(\ref{ens1}) and
$H$ into the Schr\"odinger equation, we arrive at ( setting
$\Phi(\vec{R},\vec{r},t)=
(\Phi_1(\vec{R},\vec{r},t),...,\Phi_N(\vec{R},\vec{r},t))^\top$)
\begin{eqnarray}
i\hbar\frac{\partial}{\partial t}\Phi&=&H_{eff}\Phi,\\
H_{eff}&=&\left [\frac{1}{2m}(-i\hbar\nabla_r-\vec{a})^2
+\frac{1}{2M}(-i\hbar\nabla_R-\vec{A})^2
+\epsilon+\varepsilon\right]. \nonumber \label{scheq3}
\end{eqnarray}
It is readily to show that $\vec{A}$ and $\vec{a}$ are diagonal
matrices, the elements can be expressed as
$\vec{A}_{nn}=i\hbar\langle
\alpha_n(\vec{R})|\nabla_R|\alpha_n(\vec{R})\rangle,$ and
$\vec{a}_{nn}=i\hbar\langle
\beta_n(\vec{r})|\nabla_r|\beta_n(\vec{r})\rangle.$ $\epsilon$ and
$\varepsilon$ are also $N\times N$ diagonal matrices, composed by
the eigenvalues of $H_b$ and $H_a$, respectively. In contrast with
the results presented in Eq.(\ref{scheq2}), there is no any
additional scalar potential resulting from $\vec{A}$ or $\vec{a},$
even if one of the atomic states is well separated from the
remaining atomic states. This is due to the vanishing of the
off-diagonal elements of the vector potentials $\vec{A}$ and
$\vec{a}$. The physics behind this result is the entanglement. To
show this point clearly, we suppose that the bipartite atoms are
in separable states, for example,
\begin{equation}
|\psi\rangle=\sum_n\Phi_n(\vec{R},\vec{r},t)|\alpha_n(\vec{R})\rangle_a\otimes
|\psi(\vec{r})\rangle_b,\label{sep1}
\end{equation}
where $|\psi(\vec{r})\rangle_b$ represents a state of atom b.
Clearly, for a specific $n$ that $|\alpha_n(\vec{R})\rangle_a$ is
well separated from the others, $\Phi_n$ satisfies
\begin{eqnarray}
i\hbar\frac{\partial}{\partial
t}\Phi_n=[\frac{1}{2M}(-i\hbar\nabla_R-\vec{A}_{nn})^2+V(\vec{R})]\Phi_n,
\label{scheq4}
\end{eqnarray}
with $\vec{A}$ being $N$ by $N$ matrix given by
$\vec{A}_{mn}=i\hbar\langle \alpha_m(\vec{R})|\nabla
\alpha_n(\vec{R})\rangle,$ and $V(\vec{R})=\frac{1}{2M}\sum_{j\neq
n}\vec{A}_{nj}\cdot\vec{A}_{jn}$, representing the additional
scalar potential, which appears due to the nonzero off-diagonal
elements of $\vec{A}$. The restrictions Eqs(\ref{res1}) and
({\ref{res2}) can be removed, leading to no change of the
conclusion, this can be understood as follows. Consider entangled
states of the bipartite atoms,
\begin{equation}
|\psi\rangle=\sum_n\Phi_n(\vec{R},\vec{r},t)|\xi_n(\vec{R})\rangle_a\otimes
|\gamma_n(\vec{r})\rangle_b,\label{ent3}
\end{equation}
where $|\xi_n(\vec{R})\rangle_a$ $(|\gamma_n(\vec{r})\rangle_b)$
are not the eigenstates of $H_a$ ($H_b$). In terms of the
eigenstates of $H_a$ and $H_b$,  $|\xi_n(\vec{R})\rangle_a $ and
$|\gamma_n(\vec{r})\rangle_b$ may be expressed as
\begin{eqnarray}
|\xi_n(\vec{R})\rangle_a&=&\sum_j
C_j^n(\vec{R})|\alpha_j(\vec{R})\rangle_a, \nonumber\\
|\gamma_n(\vec{r})\rangle_b &=&\sum_j
c_j^n(\vec{r})|\beta_j(\vec{r})\rangle_b.
\end{eqnarray}
The Schr\"odinger equation instead of Eq.(\ref{scheq3}) then
takes,
\begin{eqnarray}
i\hbar\frac{\partial}{\partial t}\Phi&=&
[\frac{1}{2m}(-i\hbar\nabla_r-\vec{a})^2
\nonumber\\
&+&\frac{1}{2M}(-i\hbar\nabla_R-\vec{A})^2 +e+E] \Phi,
\label{scheq4}
\end{eqnarray}
where $\vec{e}$ and $\vec{E}$ are $N$ by $N$ matrices as follows.
$e_{ij}=\langle \gamma_i|H_b|\gamma_j\rangle,$ and $E_{ij}=\langle
\xi_i|H_a|\xi_j\rangle.$ $E_{ij}$ and $e_{ij}$ would induce
population transfer, but these terms could not result in any
scalar potential even if the population transfer induced may be
ignored, because $\vec{A}$ and $\vec{a}$ are diagonal in this
case, too. This means that the atoms experience no additional
scalar potential when they evolve adiabatically in any internal
states as long as they are entangled.

 In conclusion, we have discussed the gauge fields in
ultra-cold atoms with space dependent spin-spin collisions. The
vector gauge field leads to zero magnetic fields, but gives rise
to effective potentials to the relative motion of the atoms, this
effective potential may be strong and results in escaping of the
trapped atoms from the trapping potential. For the center-of-mass
motion, the gauge field comes from relatively slow motion of the
center-of-mass, the effects of the space dependent spin-spin
collision on the center-of-mass motion take place through changing
the eigenfunction of the relative motion. It is estimated to be
large near the Feshbach resonance. The entanglement has been shown
to eliminate the additional scalar potential, explanations and
discussions are also given.

\vskip 0.3 cm We acknowledge financial support from NCET of M.O.E,
and NSF of China Project No. 10305002 and 60578014.

\end{document}